\documentstyle[epsf]{mn}

\title[{\sl XMM-Newton} EPIC observations of Her X-1]{{\sl XMM-Newton} 
EPIC observations of Her X-1\thanks{Based on observations
obtained with XMM-Newton, an ESA science mission with instruments
and contributions directly funded by ESA Member States and the USA
(NASA).}}

\author[]{Gavin Ramsay$^{1}$, Silvia Zane$^{1}$,  Mario A. 
Jimenez-Garate$^{2,4}$, 
Jan-Willem den Herder$^{3}$,\and C. J. Hailey$^{4}$ \\
$^{1}$Mullard Space Science Lab, University College London,
Holmbury St. Mary, Dorking, Surrey, RH5 6NT, UK\\
$^{2}$MIT Center for Space Research, 77 Massachusetts Avenue, Cambridge, 
MA 02139, USA\\ 
$^{3}$SRON, National Institute for Space Research, Sorbonnelaan 2,
3584 CA Utrecht, The Netherlands\\
$^{4}$Columbia Astrophysics Laboratory, Columbia University, New York,
NY 10027, USA\\}
\date{Received: } 

\begin{document}
\def\rchi{{${\chi}_{\nu}^{2}$}}
\outer\def\gtae {$\buildrel {\lower3pt\hbox{$>$}} \over
{\lower2pt\hbox{$\sim$}} $}
\outer\def\ltae {$\buildrel {\lower3pt\hbox{$<$}} \over
{\lower2pt\hbox{$\sim$}} $}                                  

\newcommand{\Msun} {$M_{\odot}$}
\newcommand{\ergscm} {ergs s$^{-1}$ cm$^{-2}$}
\newcommand{\ergss} {ergs s$^{-1}$}
\newcommand{\ergsd} {ergs s$^{-1}$ $d^{2}_{100}$}
\newcommand{\pcmsq} {cm$^{-2}$}
\newcommand{\ros} {\sl ROSAT}
\newcommand{\exo} {\sl EXOSAT}
\newcommand{\xmm} {\sl XMM-Newton}
\def\rchi{{${\chi}_{\nu}^{2}$}}
\newcommand{\Mwd} {$M_{wd}$}
\def\Mdot{\hbox{$\dot M$}}
\def\mdot{\hbox{$\dot m$}}

\maketitle

\begin{abstract}

We present spin-resolved X-ray data of the neutron star binary Her X-1
taken using the EPIC detectors on {\sl XMM-Newton}. The data were
taken at three distinct epochs through the 35 day precession
period. The energy dependent light curves of this source vary
significantly from epoch to epoch. It is known that the relative
phasing of the soft (\ltae 1~keV) and hard (\gtae 2~keV) X-rays
varies. Here, we find that the phase shift between the soft and hard
bands during the main-on state is considerably different from that
observed in the past. Further, it continues to change significantly
during the other two observations. This suggests that we are
observing, for the first time, a {\it substantial and continuous
variation in the tilt of the disk}, as it is expected if the accretion
disk is precessing in the system.  Analysis of the spin resolved data
confirms that the equivalent width variation of the fluorescence Fe K
line at $\sim$6.4~keV follows that of the soft X-ray emission in the
main-on state, thus suggesting a common origin for Fe K line and
thermal component. The Fe K line is considerably broader when the
source is brightest.

\end{abstract}


\section{Introduction}
\label{int} 

Her~X-1 is an eclipsing binary system consisting of a neutron star and
an A/F secondary star, HZ~Her.  Since its discovery, it has been
observed extensively in many wavebands revealing a high degree of
complexity in its behaviour. Her~X-1 has a spin period of $\sim
1.24$~s and a binary orbital period of 1.7~d (Tananbaum et al. 1972,
Giacconi et~al. 1973, Deeter, Boynton \& Pravdo 1981, Oosterbroek
et~al. 2001). Eclipses are seen which recur on the timescale of the
binary orbital period, indicating that the system has a high
inclination. It also varies in X-rays on a period of 35~d, with a
``main-on'' state lasting $\sim 10$~d and a secondary ``short-on''
state lasting $\sim 5$~d. Between these states the source is fainter
for $\sim 10$~d.

The origin of the 35~d cycle is thought to be due to the precession
of a tilted, warped accretion disk that periodically obscures X-rays
from the central neutron star (see e.g. Gerend \& Boynton 1976, Parmar 
et~al. 1999, Ketsaris et~al. 2000, Coburn et~al. 2000). This idea is
strengthened by the fact that optical emission from HZ~Her
persists at the same level throughout the main-on and low state,
suggesting that the companion is still being irradiated by a strong
X-ray source (Bahcall \& Bahcall 1972, Bahcall 1978, Delgado et~al.
1983).

The broadband X-ray spectrum of Her~X-1 is also extremely complex.
During the main-on state, the overall continuum is well described by a
thermal blackbody component with temperature $kT_{bb}\sim0.1$~keV,
that dominates the spectrum at low energies (McCray et~al. 1982,
Oosterbroek et~al. 1997), and a broken power-law plus an exponential
cut off at higher energies. The galactic hydrogen column density in
the direction of Her X-1 is low, $N_H \sim 10^{19}-10^{20}$~atoms
cm$^{-2}$.

In addition to these spectral features, at least three other
components have been identified: i) a feature at $\sim 1$~keV, often
referred to as the `Fe~L line' (McCray et~al. 1982, Oosterbroek
et~al. 1997; see also Mihara \& Soong 1994, Oosterbroek et~al. 2000
who reported a discrimination in two narrow lines at 0.90 and
1.06~keV); ii) a broad Fe emission feature at $\sim 6.4$~keV (Pravdo
et~al. 1977, Choi et~al. 1994, Oosterbroek et~al. 1997, Coburn
et~al. 2000); iii) a cyclotron absorption feature at $\sim
40$~keV. The latter gives an inferred value of the magnetic field
$\sim 3\times 10^{12}$~G (see e.g. Dal Fiume et~al. 1997).

An analysis of {\sl Ginga}, {\sl RXTE} and {\sl BeppoSax} data (Mihara
et~al. 1991, Oosterbroek et~al. 1997, Coburn et~al. 2000, Oosterbroek
et~al. 2000) has shown that out with the main-on state the effects of a
significant intervening absorption (with typical $N_H$ $> 10^{21}$
atoms cm$^{-2}$) are clearly seen as a change in the spectral
shape. However, a substantial flux remains below 0.5~keV that should
be completely absorbed by such material unless the associated covering
factor is partial (Oosterbroek et~al. 2000).  There are two
possibilities which cannot be spectrally distinguished: a) the
presence of two separate ``scattering'' and ``absorbed'' regions; b)
an intrinsic partial covering of the emitting region, i.e. a situation
in which the emission spectrum is the same in the low and main-on
states, but a fraction of the emerging radiation is highly scattered
and absorbed. Both cases are referred in the literature as ``partial
covering'' models, and b) is what is expected, for instance, during
occultation phases of the neutron star by the accretion disk.

Another manifestation of the 35~d precession cycle is the repeating,
systematic evolution of the X-ray pulse profile over this cycle.
Extensive coverage of these variations have been obtained using
$Ginga$ (Deeter et~al. 1998). Simultaneous X-ray and UV observations
have also been carried using $ASCA$ and $HST$ by Boroson et
al. (1996), showing that the pulse shape becomes smooth and sinusoidal
from the soft X-ray band into the UV (see also Leahy, Marshall, \&
Scott 2000). Past attempts to model the pulse changes in various
energy bands relied on a combination of free neutron star precession
and disk obscuration, as well as obscuration by flaps of matter at the
magnetosphere or obscuration by the tilted disk (see Deeter
et~al. 1998, Scott, Leahy \& Wilson 2000).  However, most of these
studies fail in explaining more than a few aspects of the complex
pulse evolution. Recently, detailed interpretations have been
presented by Deeter et~al. (1998) and Scott et~al. (2000). In
particular, the model by Scott et~al. (2000) is based on inner disk
occultation of the X-ray beam from the neutron star. The latter, in
turn, consists of two components: a pencil direct beam and a fan beam
emission in the antipodal direction (see \S \ref{enres}). This model
refines the disk and pulsar beam geometry and qualitatively accounts
for {\it both} the pulse phase {\it and} its evolution during the
main-on and short-on high states.
 
In this paper we present the results of a series of three observations
made using the EPIC cameras on board {\sl XMM-Newton} at different
$\Phi_{35}$ (see \S \ref{obs}). We discuss the spin period of the
neutron star in \S \ref{spin} and the pulse profile in different
energy bands are reported in \S \ref{pulse}. In \S \ref{spec} we
report the pulse-averaged broadband spectra, and in \S \ref{pulsespec}
we examine the pulse resolved spectra. We discuss our results in \S
\ref{disc}.

\section{Observations}
\label{obs}

Her X-1 was observed using {\sl XMM-Newton} on 3 separate occasions in
2001, the details of which are summarized in Table~\ref{obslog}. The
EPIC detectors (Turner et~al. 2001; Str\"{u}der et~al. 2001), were
configured in timing mode (PN and MOS1) and full frame mode (MOS2) for
all observations. The MOS2 data were heavily piled up and therefore
are not used in the analysis. The medium filter was used in all
observations. The {\xmm} RGS observations are presented in a separate
paper by Jimenez-Garate et~al. (2002).

\begin{figure}
\begin{center}
\setlength{\unitlength}{1cm}
\begin{picture}(8,5)
\put(-1.5,-11.5){\includegraphics{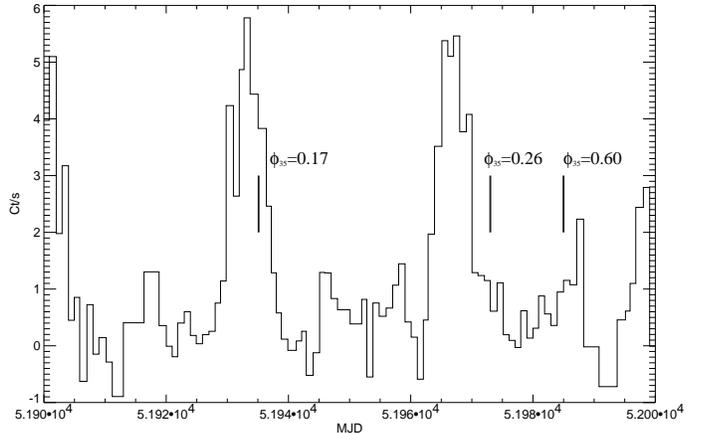}}
\end{picture}
\end{center}
\caption{The {\sl RXTE/ASM} (2-10)~keV light curve of
Her~X-1:
the thick lines indicate the {\sl XMM-Newton} observations.}
\label{asm} 
\end{figure}

To determine how these observations relate to the 35~d precession
period, we extracted the {\sl RXTE ASM} (2-10)~keV quick-look light
curve (Figure~\ref{asm}). This shows that the first observation took
place close to maximum X-ray brightness, while the third was close to
the secondary maximum. The second observation took place after the end
of the main-on state. The values of $\Phi_{35}$ corresponding to the
three datasets are also reported in Table~\ref{obslog}, where
$\Phi_{35}$=0.0 is the main-on state turn-on.

\begin{table}
\begin{tabular}{llllr}
\hline
\small
Observation& Effective&EPIC PN &Orbital &35~d\\
Date (orbit) & Exposure&Mean ct&Phase&Phase\\
     &  (ksec)  & (ct/s)&     & \\
\hline
2001 Jan 26 (207) & 10& 398.4 &0.19--0.25&0.17\\
2001 Mar 4 (226) & 11& 21.4 &0.46--0.55&0.26\\
2001 Mar 17 (232) & 11 & 54.3 & 0.51--0.59&0.60\\
\hline
\end{tabular}
\caption{Summary of the {\sl XMM-Newton}
observations of Her X-1. For the orbital phasing we use the ephemeris
of Still et al. (2001); for the 35~d phasing we define phase 0.0 at the
main high state turn-on.}
\label{obslog}
\end{table}
\normalsize

Data were processed using version 5.3.3 of the {\sl XMM-Newton}
Science Analysis System. Because the point spread function of the
source covers most of the CCD which is read out, we are not able to
reliably extract background light curves or spectra in the timing mode
data.  We did however examine the background light curves in the full
frame data taken using the MOS2 detector. The background was
relatively constant and low in the first (orbit 207) and last (orbit
232) observations. However, in the second observation (orbit 226) the
particle background was high near the start and end of the
observation: these times were excluded in the analysis. Since Her X-1
is much brighter than the background, we do not expect that the
absence of background data will have a significant effect on our
results.

We also had to correct two of the EPIC pn datasets for time jumps. In
the case of orbit 207 a 1 sec time jump was found to take place at
frame numbers 286698 and 646626. The cause of these time jumps is not
known but are easily corrected for by subtracting 1 sec from the frame
times for events occurring after these frame numbers. We also had to
correct orbit 232 at frame number 514275 where {\tt FTCOARSE} was
incremented by 1 sec when it should not have done.

In this paper, in the interest of brevity, we report only the results
derived using the EPIC pn data. This was because this data had a
higher signal to noise than the EPIC MOS data. However, the results
derived from the MOS data are consistent with those derived from the
EPIC pn data.

This paper refines the work of Ramsay et al (2002) who present a
prelimenary report of these {\xmm} observations.  In particular, the
data reported by Ramsay et al were folded on different periods to the
values reported below. This is because the routine used to correct
events to the solar system barycenter ({\tt barycen v1.9}) was
incorrect. The spin folded light curves for $\Phi_{35}$=0.26 and 0.60
reported by Ramsay et al (2002) are very similar to our present
findings. In the case of the $\Phi_{35}$=0.17 data, the general shape
of the light curves are similar, but the details differ.

\section{The spin period}
\label{spin}

The spin period of Her X-1 has been found to vary from $\sim$1.23772~s
to 1.23782~s, with alternating phases of spin-up and spin-down (Parmar
et~al. 1999, Oosterbroek et~al. 2001). Furthermore, it can vary
significantly on relatively short timescales. Before determining the
spin period from our data, we first performed a barycentric correction
(using {\tt barycen v1.13.2}) and also a correction for the motion
around the binary center of mass (using the ephemeris of Still
et~al. 2001) on each event photon (this was after the correction for
the 1 sec time jumps reported above). We then determined the period,
analyzing the data from each epoch separately using a Discrete Fourier
Transform (Deeming 1975, Kurtz 1985). The best fit periods determined
from the EPIC MOS and pn detectors are reported in Table~\ref{period}.

The spin period at each epoch is consistent between the two EPIC
detectors. Further the spin period taken from the first epoch (2001
Jan 26) is continuing to slow down after its long anomalous slow
state: Oosterbroek et al (2001) determined a spin period of
1.2377697(3) in Oct 2000 using {\sl BeppoSAX} data. Her X-1 was
observed on 2001 Jan 25 using {\sl Chandra}. The spin period was found
to be 1.237786 sec (Burwitz, priv. comm). This spin period is
marginally slower compared to our observation of Jan 26. Since the
{\sl Chandra} observation was configured in Grating mode and the frame
time is slower than the EPIC timing mode, the {\xmm} spin period is
likely to be more accurate. There is a slight decrease in the spin
period recorded in the two following epochs (2001 March).

\section{Spin resolved light curves}
\label{pulse} 

\begin{figure*}
\begin{center}
\setlength{\unitlength}{1cm}
\begin{picture}(6,10.5)
\put(14,-2.2){\includegraphics{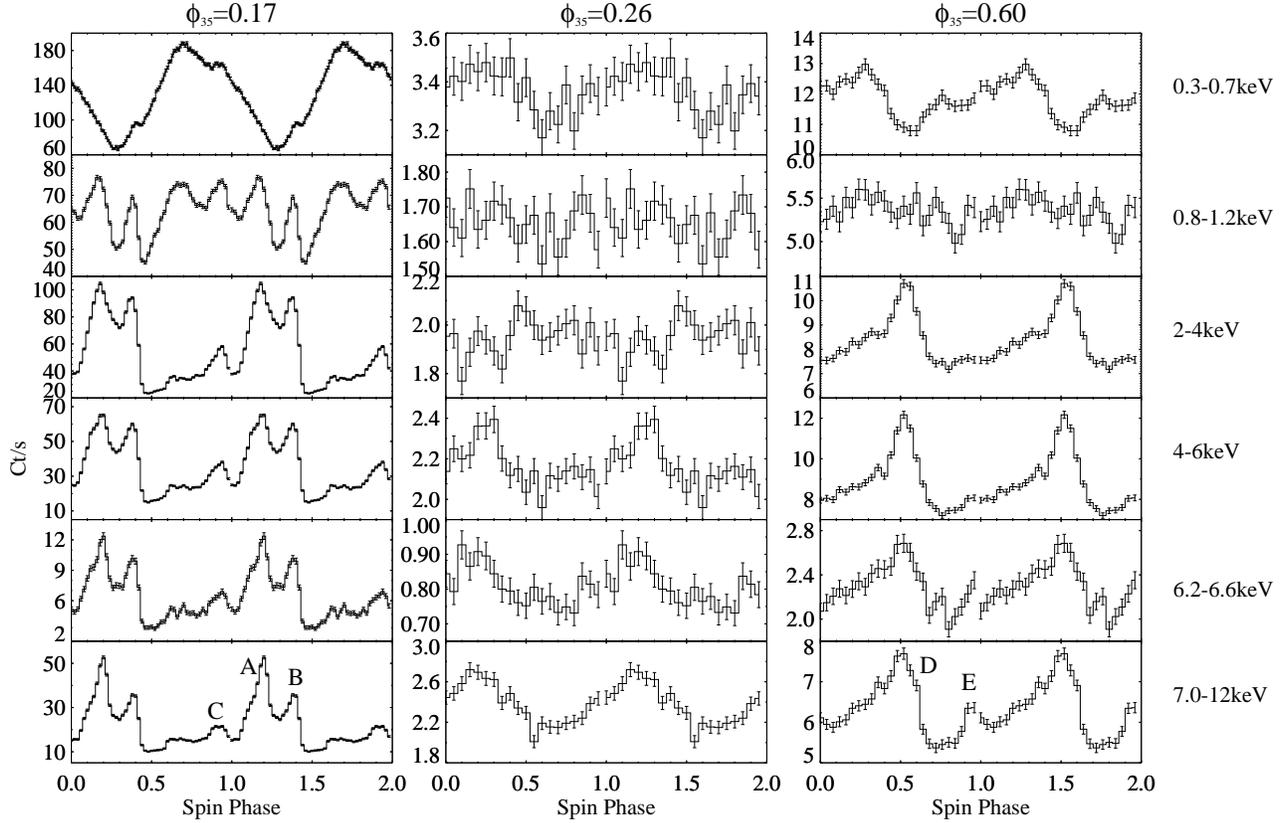}}
\end{picture}
\end{center}
\caption{The spin profiles for various energy bands for the three
epochs. Because of the uncertainty in the spin period in each epoch,
the spin phases are not on the same absolute scale. The captions in
the 7--12 keV band panels are refered to in the text.}
\label{fold} 
\end{figure*}

We have extracted the light curves in various energy bands, at each of
the three epochs. We then performed a least squares sinusoid fit to
the light curves, using the best fit period appropriate to the epoch
as previously found. The pulsed fraction, defined here as the
modulation amplitude in the various energy bands divided by the mean
count rate in that band, is shown in Table~\ref{amptable}. We also
folded and binned the same light curves, which are shown in
Figure~\ref{fold}.

\begin{table}
\begin{tabular}{lrr}
\hline
Observation& EPIC MOS & EPIC pn\\
 & (s) \\
\hline
2001 Jan 26& 1.2377727(7) & 1.237774(5)\\
2001 Mar 4& 1.23779(2) & 1.237753(6)\\
2001 Mar 17& 1.2377594$^{+13}_{-5}$ & 1.237751(3)\\
\hline
\end{tabular}
\caption{The best fit periods determined from the EPIC MOS and pn
light curves using a Discrete Fourier Transform. The number in
parenthesis is the error on the last digit.}
\label{period}
\end{table}

\begin{table*}
\begin{tabular}{lllllll}
\hline
$\Phi_{35}$&0.3--0.7&0.8--1.2&2--4&4--6&6.15--6.7&7.0--12.0\\
           &keV&keV&keV&keV&keV&keV\\
\hline
0.17 & 0.399$\pm$0.004& 0.097$\pm$0.003 & 0.400$\pm$0.004 & 
0.292$\pm$0.004 & 0.341$\pm$0.012 & 0.235$\pm$0.004\\
0.26 & 0.029$\pm$0.005 & 0.022$\pm$0.010 & 0.027$\pm$0.010 & 
0.049$\pm$0.009 & 0.071$\pm$0.013 & 0.111$\pm$0.009\\
0.60 & 0.059$\pm$0.007& 0.024$\pm$0.008 & 0.136$\pm$0.042 & 
0.164$\pm$0.006 & 0.098$\pm$0.009 & 0.110$\pm$0.006\\
\hline
\end{tabular}
\caption{The pulsed fraction (modulation amplitude/mean count rate) over
various spectral ranges for the 3 epochs.}
\label{amptable}
\end{table*}

As we can see from Table~\ref{amptable}, the relative modulation
amplitude is strongest at $\Phi_{35}=0.17$ (i.e. close to the main-on
state). At this epoch, the value of the pulsed fraction is similar in
all energy bands with the exception of the (0.8-1.2)~keV range in
which it is significantly lower. At $\Phi_{35}=0.26$, the relative
modulation amplitude is strongly reduced at all energies, but
increases towards higher energies. At $\Phi_{35}=0.60$, the relative
modulation amplitude is again similar at energies above 2~keV and is a
factor $\sim 2-3$ lower compared to that found in $\Phi_{35}=0.17$.

The folded light curves corresponding to the different epochs are
dramatically different (Figure~\ref{fold}). The softest X-ray emission
(0.3-0.7keV) always shows a smooth, almost sinusoidal shape that is
most prominent at $\Phi_{35}=0.17$.  At this epoch, the hard emission
shows a prominent double peaked maximum, with the `trailing' peak
(peak `B' in Figure~\ref{fold}) being less intense especially at
harder energies. The phasing of those peaks corresponds to the minimum
in the (0.3-0.7~keV) band. The (0.8-1.2~keV) light curve shows both
the double peak seen at higher energies and also the broad maximum
seen at lower energies. The (0.8-1.2)~keV band shows a superposition
of both the soft X-ray component and also the hard X-ray component.

In contrast, in the low-state ($\Phi_{35}=0.26$) epoch the soft and
hard X-ray light curves are very similar in shape and are roughly in
phase. 

Close to the short-on state ($\Phi_{35}=0.60$), the maximum of the
hard X-ray light curve corresponds to the minimum in the soft X-ray
light curve. Further, at higher energies a secondary peak becomes more
prominent (peak `E' in Figure \ref{fold}).

We also show the light curves in the energy range (6.2-6.6)~keV.  This
energy range includes the emission from the Iron K$\alpha$ emission
line. In the main-on and also the low state, the folded light curve in
this energy band is very similar to that of the (7-12)~keV folded
light curve. In the short-on state the general shape is very similar,
although shortly after the descent from spin maximum there is a rise
in flux compared to the minimum seen at higher energies. It is
difficult to isolate the line emission from the continuum emission
using these kind of light curves. We examine this further in \S
\ref{pulsespec}, where we present the pulse resolved spectra.

\begin{figure}
\begin{center}
\setlength{\unitlength}{1cm}
\begin{picture}(12,5.5)
\put(-0.5,0.0){\includegraphics{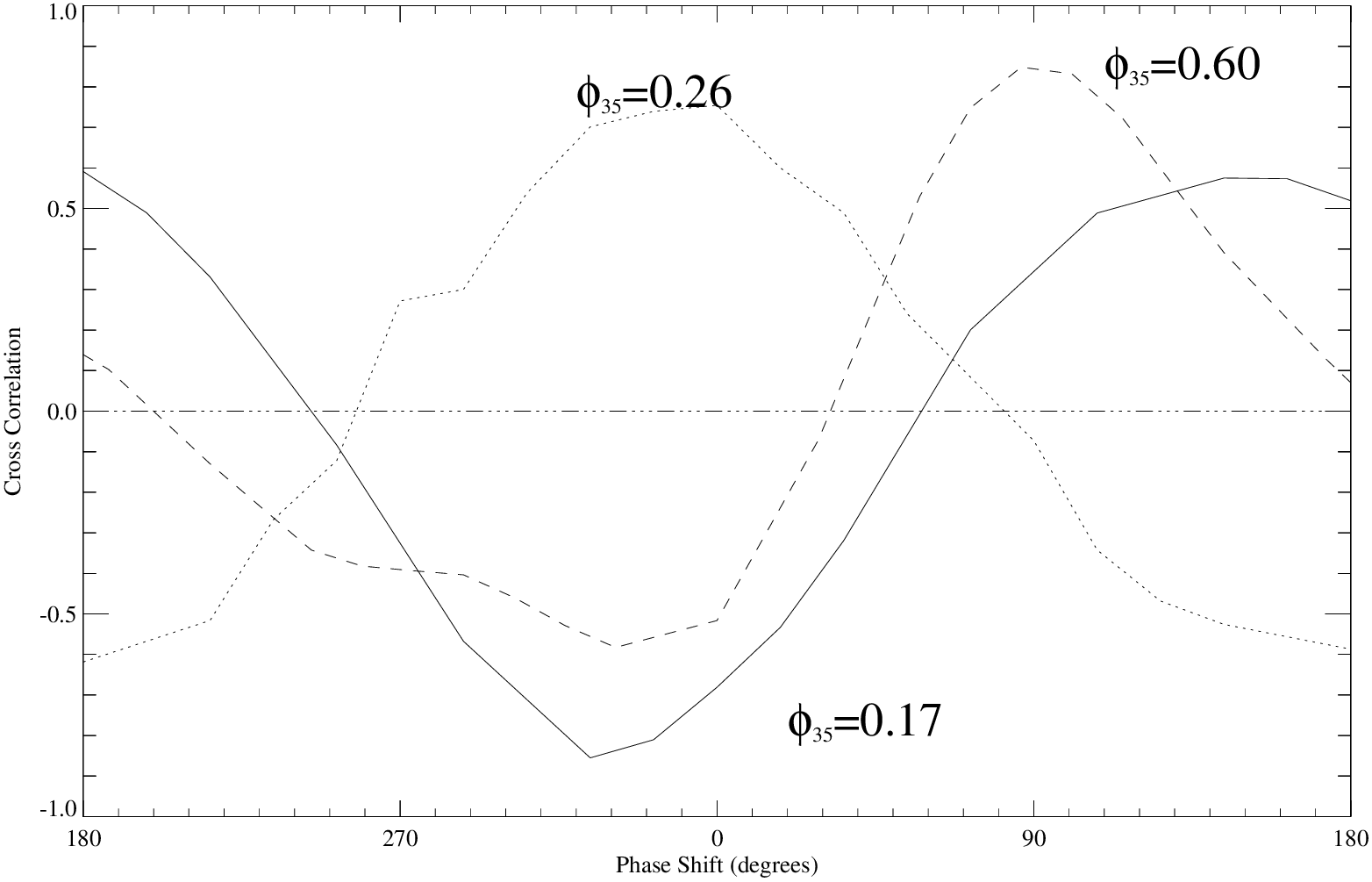}}
\end{picture}
\end{center}
\caption{The cross correlations for the soft and hard energy resolved
light curves at the 3 epochs.}
\label{crosscor} 
\end{figure}

Comparing the relative phasing of the soft and hard band we find that
a turnover is evident after $\sim 2$~keV, in the main-on and short-on
states. To determine the shift in phase between the hard and soft
emission at each epoch, we have cross correlated the (0.3--0.7)~keV
and the (2--4)~keV light curves for the two states $\Phi_{35}=0.17$
and $\Phi_{35}=0.60$. For the fainter state ($\Phi_{35}=0.26$) we
determined the cross correlation between the (0.3--0.7)~keV and
(7--12)~keV light curve, since the modulation in the latter band was
higher than in the (2--4)~keV band. We show the cross correlation in
Figure \ref{crosscor}.

Considering the main-on state first, we find that the most prominent
cross correlation peak shows the soft and the hard light curves are
strongly anti-correlated, with only a small phase lag ($\sim
330^\circ$). We also find a strong positive peak at $\sim 150^\circ$,
which corresponds to the separation between the soft X-ray peak and
the general peak in hard X-rays. The faint state shows a negative
correlation at $\sim 180^\circ$ and a positive correlation at small
phase lag ($\sim 350^\circ$), with the positive correlation having a
higher coefficient. Again, the positive peak corresponds to the
separation between maxima in the soft and hard lightcurves -- they are
basically in phase. During the short-on state, we find the highest,
positive correlation near $\sim 90^\circ$, which is due to the shift
between the maximum in the soft and hard X-rays. It is evident that
the relative phase shift between the soft and hard X-ray light curves
varies as a function of the 35~d precession period. We discuss this
further in \S \ref{disc}.

\section{The pulse averaged X-ray Spectra}
\label{spec} 

\begin{figure}
\begin{center}
\setlength{\unitlength}{1cm}
\begin{picture}(12,5.5)
\put(-0.8,-0.5){\includegraphics{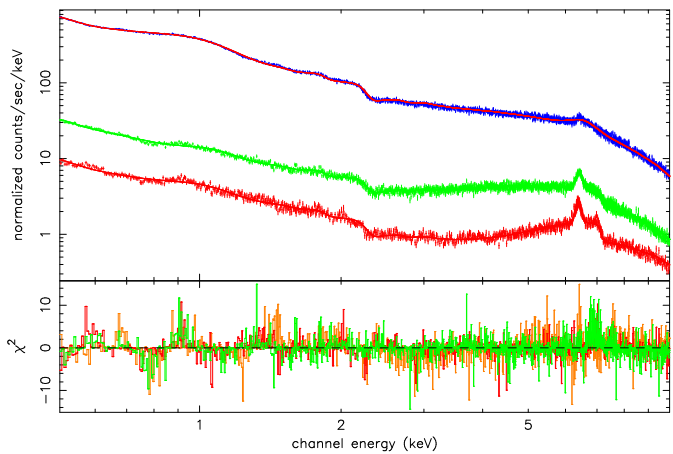}}
\end{picture}
\end{center}
\caption{The EPIC pn spectra of Her X-1. From the top
to the bottom: the spectra taken on Jan 26 ($\Phi$=0.17), March 17 
($\Phi$=0.26) and March 4 ($\Phi$=0.60).}
\label{intspec} 
\end{figure}

As recommended by the {\xmm} Science Operation Center, we ignore
energies less than 0.5 keV in our spectral fits. The energy response
file we use (epn$\_$ti40$\_$sdY9$\_$medium.rsp), assumes only single
and double pixel events, therefore we extracted only these events to
construct the EPIC pn spectra.  We extracted pulse averaged spectra at
each epoch.  We show the pulse averaged spectra taken at the three
$\Phi_{35}$ in Figure~\ref{intspec}. The immediate points to note are
the apparent hardening of the spectra at lower intensity and the Fe K
fluorescent line near 6.4~keV, that appears to become more prominent
as the intensity weakens.

To model these spectra, we used an absorbed power law together with a
blackbody and a Gaussian near 6.4~keV to account for the fluorescence
line. For the absorption model, we included a column of neutral
Hydrogen fixed at the low value of 5$\times10^{19}$ cm$^{-2}$, which
accounted for the interstellar absorption towards Her~X-1. If a good
fit was still not achieved we added additional partial covering models
of neutral Hydrogen. We also included an additional Gaussian line at
$\sim$1 keV which was used to model the excess of emission at these
energies found in the RGS spectra of the main-on state (Jimenez-Garate
et~al. 2002).

\small
\begin{table*}
\begin{center}
\begin{tabular}{lrrr}
\hline & $\phi_{35}$=0.17 & $\phi_{35}$=0.26 &
$\phi_{35}$=0.60\\ 
\hline 
$N_{H}$ (atoms cm$^{-2}$), cvf & &
56.1$^{+0.7}_{-0.7}\times10^{22}$, 0.79$\pm$0.01&
22.0$\pm0.5\times10^{22}$, 0.54$\pm0.05$\\ 
 $N_{H}$ (atoms cm$^{-2}$), cvf & &
10.3$^{+1.2}_{-0.7}\times10^{22}$, 0.29$\pm$0.01 &
5.7$\pm0.1\times10^{22}$, 0.51$\pm0.01$ \\ 
$kT_{bb}$ (eV) & 101$\pm$1 & 104$^{+1}_{-2}$ & 102$\pm1$\\ 
$\alpha$ & 0.90$\pm$0.01 & 0.87$^{+0.05}_{-0.01}$ &
1.05$\pm0.01$ \\ 
Fe K line center (keV)& 6.540$^{+0.018}_{-0.017}$ &
6.385$^{+0.005}_{-0.007}$ & 6.417$\pm0.005$\\ 
FWHM (keV) & 0.28$^{+0.03}_{-0.02}$ & 0.059$_{-0.013}^{+0.006}$ &
0.069$^{+0.008}_{-0.007}$\\ 
line normalisation & 4.9$\pm0.4\times10^{-3}$ &
7.0$^{+0.3}_{-0.5}\times10^{-4}$&
9.8$^{+0.24}_{-0.3}\times10^{-4}$\\ 
EW (eV) & 164$^{+13}_{-12}$ & 208$_{-14}^{+8}$ & 179$^{+10}_{-6}$ \\ 
Observed flux (0.3-10)~keV &
3.38$\pm0.01\times10^{-9}$ & 1.10$\pm0.03\times10^{-10}$&
3.35$\pm0.01\times10^{-10}$ \\ 
Blackbody observed flux (0.3-10)~keV
& 3.53$\pm0.05\times10^{-10}$& 4.21$\pm0.13\times10^{-12}$&
1.32$\pm0.15\times10^{-11}$\\ 
Blackbody bolometric flux &
6.62$\pm0.05\times10^{-10}$ & 5.18$\pm0.15\times10^{-11}$ &
1.06$\pm0.02\times10^{-10}$\\ 
Power law observed flux (0.3-10)~keV &
2.82$\pm0.05\times10^{-9}$& 1.02$\pm0.11\times10^{-10}$ &
3.12$\pm0.02\times10^{-10}$\\ 
Power law unabsorbed flux (0.01--99) keV &
3.65$\pm0.05\times10^{-8}$& 4.3$\pm0.1\times10^{-9}$ &
4.95$\pm0.02\times10^{-9}$\\ 
\rchi & 1.35 (1685) & 1.18 (943) & 1.31 (1673) \\ 
\hline
\end{tabular}
\end{center}
\caption{The best fit parameters to the EPIC pn pulse averaged spectra. All
errors are quoted at the 90 percent level. 
A constant column density of 5$\times10^{19}$
atoms cm$^{-2}$ was also included, to account for the interstellar
absorption. The units for the fluxes are
erg~cm$^{-2}$~s$^{-1}$; the line normalisation is in
photons~cm$^{-2}$~s$^{-1}$; FWHM is the full width half maximum; EW is
the equivalent width and ``cvf'' is the covering factor.}
\label{bestfits}
\end{table*}
\normalsize

\begin{table}
\begin{center}
\scriptsize   
\begin{tabular}{rrrr}
\hline
\scriptsize  
&$\Phi_{35}$=0.17 & $\Phi_{35}$=0.26 & $\Phi_{35}$=0.60 \\
\hline
 \rchi & 1.29 (1205) & 1.09 (547) & 1.17 (1195) \\
 $\alpha$ & 0.85 & fixed & fixed \\
  PL Norm & 0.148 & fixed & fixed \\ 
 $N_{H}$, cvf & & 52$^{+2}_{-3}$, 0.82$\pm0.01$ & 12.6$\pm0.5$, 0.62$\pm0.01$\\
 $N_{H}$, cvf &  & 1000${-50}$, 0.89$ \pm 0.01$ & 1500$_{-1000}$, 0.85$\pm0.01$\\
\hline
\end{tabular}
\end{center}
\caption{The best fits to the EPIC pn spectra assuming a constant
(power law) emission model based on the data obtained at
$\Phi_{35}$=0.17. We fit data in the energy bands (2.0-6.1)~keV and
(7-10)~keV. The power law (PL) normalisation is in
photons~cm$^{-2}$~s$^{-1}$~keV$^{-1}$ (at 1~keV), $N_{H}$ is in units
of $10^{22}$ atoms~cm$^{-2}$, and ``cvf'' is the covering factor. When
we allow the slope and normalisation of the power law to vary the
goodness of fit is not significantly different.}
\label{constant}
\end{table}

\normalsize

We show the spectral fits to the data in Table~\ref{bestfits}, where
we also report the total observed flux in the (0.3-10)~keV band, as
well as the bolometric, unabsorbed blackbody flux and the unabsorbed
power law flux. The fits to all three integrated spectra are reasonably
good (\rchi $<$ 1.4).

We find that the absorption is highest at $\Phi_{35}$=0.26, (when the
source was at lowest intensity), and lowest at $\Phi_{35}$=0.17 (when
the source was at highest intensity). We also find that we need to
include a blackbody component whose temperature is $\sim$0.10 keV at
each epoch. The fluorescence line at $\sim 6.4$~keV probably
originates from irradiation of cold material (see \S \ref{pulsespec}
and \S \ref{line}). We find that its flux is greatest at
$\Phi_{35}$=0.17, when the line energy is higher compared to the other
epochs. In addition, the line broadening changes significantly during
the 35~d cycle: the full width half maximum (FWHM) increases by a
factor $\sim$4 when the source is brighter. The equivalent width is
significant higher in the low state compared to the main-on or
short-on state.  We also searched for a second line at higher energies
(features at 6.4~keV and 6.7~keV have been resolved with ASCA, Endo
et~al. 2000), but there was no evidence for a second emission
feature. We discuss the spin variation in the Fe line profile in the
next section.

The ratio of the unabsorbed, bolometric blackbody luminosity to the
unabsorbed power law luminosity in the 0.01-100 keV range is
$\sim$1--2 per cent in all three epochs. This is significantly lower
than that found by Endo et al (2000) using {\sl ASCA} data (who found
a ratio of 0.11)

In order to determine if the same (power law) model for the hard
emission satisfactorily fits (apart from a local absorber) the spectra
taken at the different epochs, we first re-computed the best-fit
parameters of the power law for the $\Phi_{35}$=0.17 spectrum. We
ignored energies less than 2keV (where the blackbody component become
visible and absorption becomes increasingly important) and energies
between (6.1-7.0)~keV (where the 6.4~keV fluorescence line is
prominent). Keeping this model fixed, we then fitted the spectra taken
at $\Phi_{35}$=0.26 and $\Phi_{35}$=0.60, by allowing the absorption
to vary (adding a partial covering model if appropriate). We show the
goodness of fits in Table~\ref{constant}.  We find that a single power
law emission model is consistent for all three epochs and the
variation in the X-ray spectra is due to a variation in the amount of
absorption. In a second step, we allowed the power law parameters to
vary: the goodness of fit was not significantly different.

\section{Pulse-resolved spectroscopy}
\label{pulsespec} 

\begin{figure*}
\begin{center}
\setlength{\unitlength}{1cm}
\begin{picture}(14,11)
\put(-3.,-0.8){\includegraphics{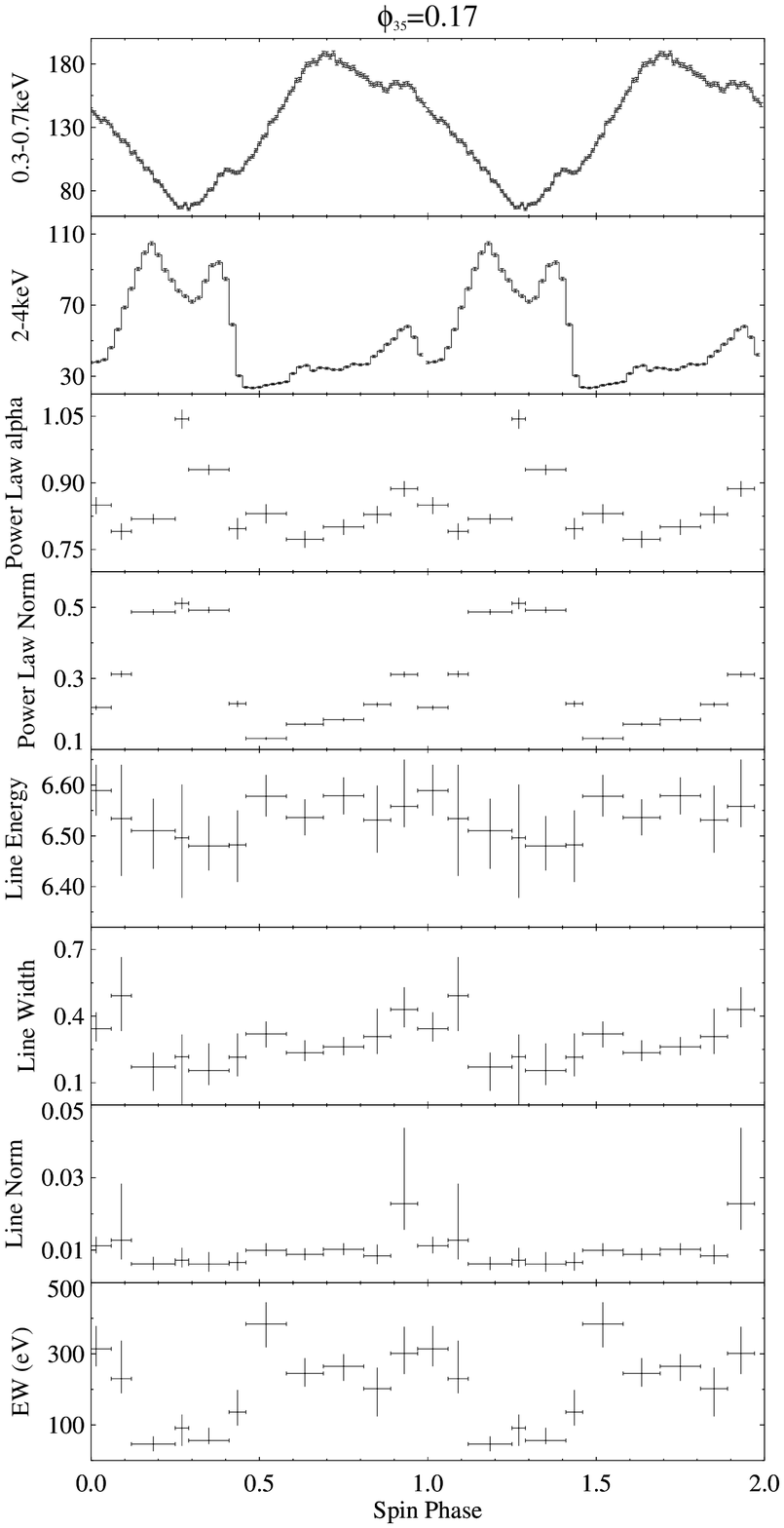}}
\put(3.,-0.8){\includegraphics{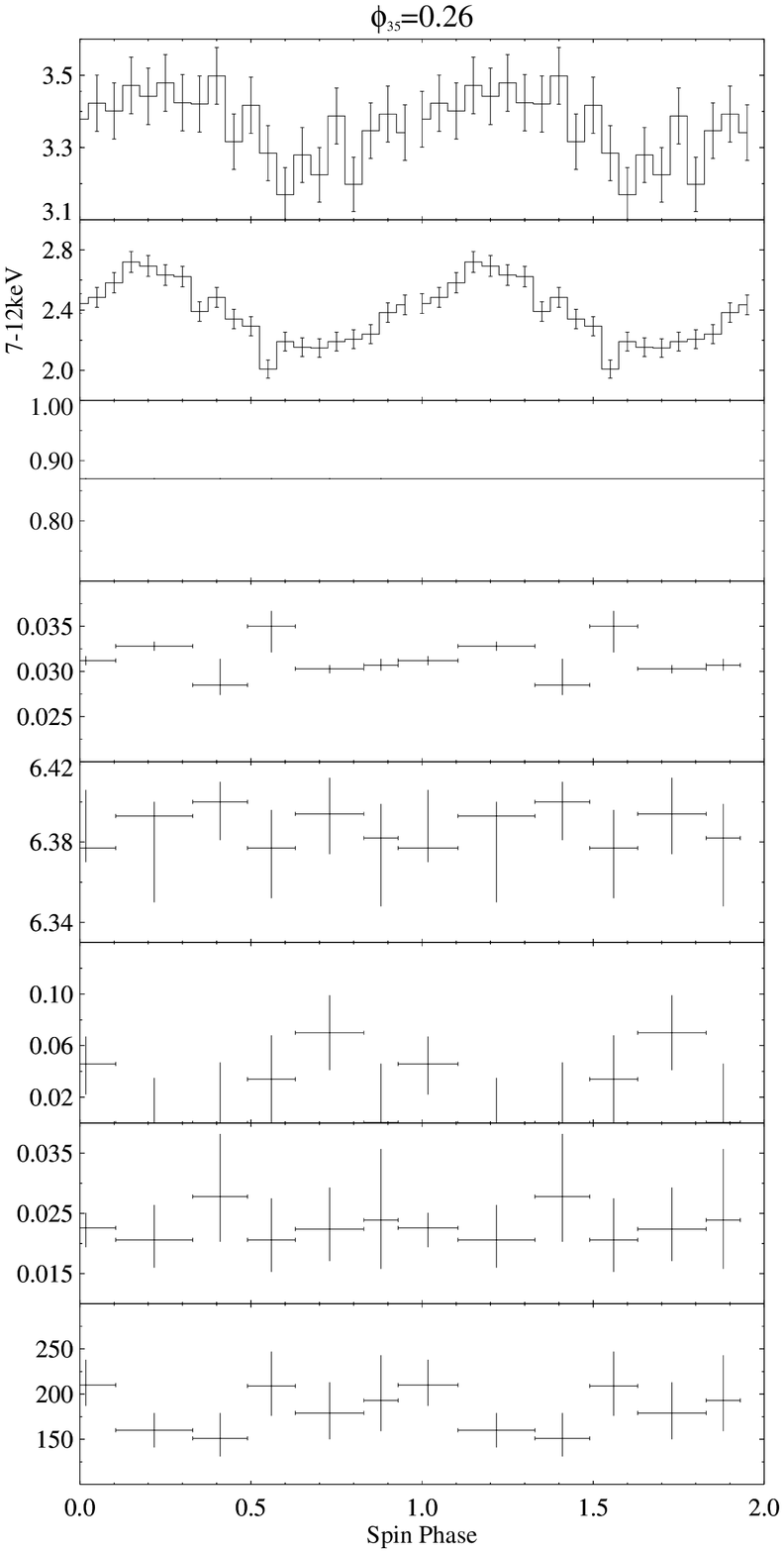}}
\put(9.2,-0.8){\includegraphics{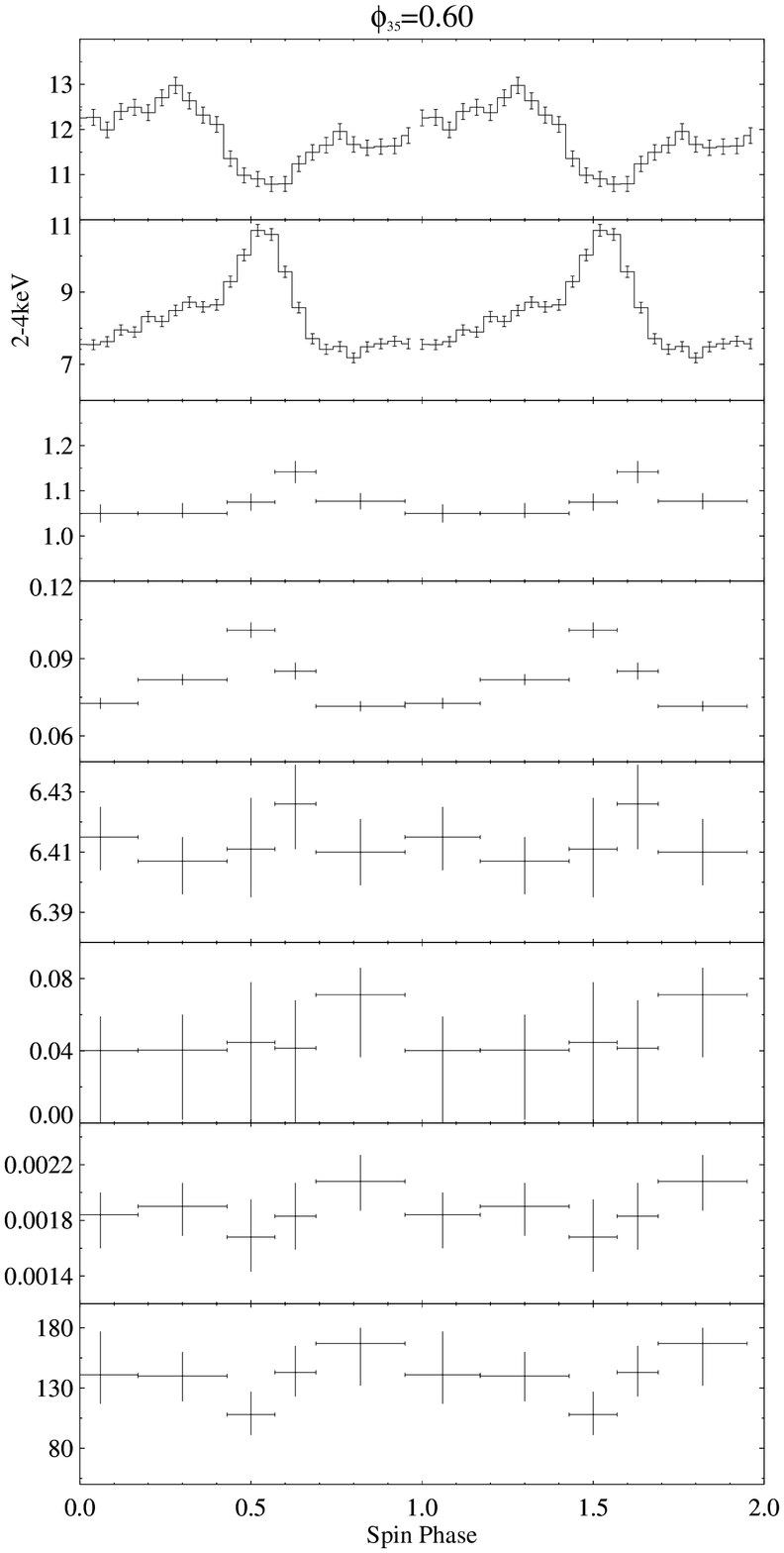}}
\end{picture}
\end{center}
\caption{The variation of the Fe line parameters as a function of spin
phase, for the three different epochs $\Phi_{35}$. We also show the
normalization of the power law component and the intensity curves in
the soft (0.3-0.7~keV) and hard energy band.  The latter is taken to
be (2-4)~keV for $\Phi_{35}$=0.17, $\Phi_{35}$=0.60 and (7-12)~keV for
$\Phi_{35}$=0.26. The line width is the FWHM. The slope of the power
law was fixed at $\Phi_{35}$=0.26.}
\label{fitph}
\end{figure*}

In order to investigate how the profile of the Fe line at 6.4~keV
varies over the $\sim$1.24~s spin period, we extracted spin resolved
spectra from the EPIC data. We ignore energies below 5~keV and above
8~keV, and we used a Gaussian plus power law model to fit the
spectra. We fix the absorption at the value found from the pulse
averaged spectra (cf \S \ref{spec}). We also fitted an absorbed power
law over the range (2.0-6.1)~keV and (7-10)~keV -- again we fix the
absorption. The results are shown in Figure~\ref{fitph}, together with
the soft and hard light curves. (In the case of the $\Phi_{35}$=0.26
data we also fix the slope of the power law at the best fit value
determined from the pulse averaged spectra since these data had the
lowest signal to noise).

At $\Phi_{35}$=0.26 and 0.60 there is no significant variation in the
line parameters. At $\Phi_{35}$=0.17, where the data have the highest
signal to noise ratio, there is clear evidence that the equivalent
width varies over the spin period with the minimum occurring at the
phase corresponding to the maximum in hard X-ray intensity. Conversely
the equivalent width is at maximum at the same spin phase that the
soft X-ray is at its broad maximum intensity. This further supports
the view that the 6.4~keV fluorescence Fe line and the black body
component originate from a common region. There is some evidence that
the line width varies over the spin phase with a maximum occurring
immediately before the rise to hard X-ray maximum.

As expected the normalisation of the power law component follows the
hard X-ray intensity modulation in the main-on and secondary-on
state. In the case of the low state there is no evidence for a
significant variation.

\section{Discussion} 
\label{disc}

The results reported in this paper show for the first time that there
is a continuous variation in the relative phase offset between the
soft and hard X-ray components. We also find that normalisation and
equivalent width of the fluorescence Fe line at 6.4~keV are modulated
in phase with the soft X-ray component. Further, we find that for
energies above $\sim$2~keV the three spectra corresponding to three
different $\Phi_{35}$ are consistent with a single emission model and
an absorption component which is lowest at $\Phi_{35}$=0.17, and
highest at $\Phi_{35}$=0.26. Both the inferred values of $N_H$ and
covering factor for the intervening matter are consistent with Coburn
et~al. (2000), with the former considerably higher than that reported
by Oosterbroek et~al. (2000). We also require more than one partial
covering component, and both these facts hint towards a fairly complex
substructure of the thermal emission. On the other hand, the nature of
the soft X-ray emission is not well understood and deserves a further,
more detailed investigation. We now proceed by discussing the physical
implications of our results.

\subsection{The energy resolved light curves}
\label{enres} 

The pulse profile of Her~X-1 is known to evolve through the main-on
state; those reported here closely resemble that observed using
$Ginga$ (Deeter et~al.  1998). The EPIC observations taken at
$\Phi_{35}=0.17$ and $\Phi_{35}=0.26$ can be qualitatively compared to
Figure~6 of Deeter et~al. (1998; main state D), while those taken at
$\Phi_{35}=0.60$ are similar to their Figure~7 (short state C). Many
of those features are naturally explained within the scenario proposed
by Scott et~al. (2000). This model was originally developed to explain
$Ginga$ and $RXTE$ observations, and it is based on the obscuration of
a multi-component X-ray beam by a counter-precessing, tilted, twisted
disk.  For simplicity, the X-ray beam is assumed to be decoupled from
the disk and is axi-symmetric; the observer's line of sight must be
close to the binary plane (to explain two maxima per 35~d cycle). One
of the main features of this model is that it ascribes the variations
observed in the pulse profile over the 35~d cycle to occultation from
the {\it inner} part of the disk, whereas most of the previous
investigations have assumed an occultation from the {\it outer}
boundary (but see also Petterson, Rothschild \& Gruber, 1991, who
argued that vertical ``flaps'' which form near the magneto-spheric
radius are partly responsible for the 35~d phase behaviour).  Inner
disk phenomena, in fact, are more plausible in explaining the peaks
evolution of the spin profile, as well the differences in the
pulse-shape of main-on and short-on. Such differences require that the
size of the dominant pulse-emitting region is roughly the scale-height
of the inner disk, so that the two high states are caused by
progressive occultation of an {\it extended} source.

The complex pattern of peaks observed in Her X-1 can be explained in
terms of successive occultations of a beam consisting of a direct pencil
beam (that originates close to the poles at the surface of the neutron
star) plus a reversed fan beam that is focused in the antipodal
direction. These two components are superimposed on three other
constant flux components: two due to magneto-spheric emission and one
due to low-state coronal emission. In the ``reversed fan beam''
geometry, the fan beam component originates at the same distance above
the neutron star surface and has an opening angle $> 90^\circ$. A
similar configuration has been predicted theoretically by Brainerd \&
Meszaros (1991), who studied the backscattering of the magnetic polar
cap radiation by the incoming accretion flow and its subsequent
gravitational focusing around the neutron star. The accretion column
is predicted to be optically thin to Thomson scattering, while the
fan beam photons are produced by cyclotron resonance scattering.

The overall situation is summarized in the bottom panel of Figure 8 in
Scott et~al. (2000), while their Figures 10a-10b illustrate the
evolution of the pulse profiles predicted during the main-on and
short-on respectively.  A number of the features observed in our {\sl
XMM-Newton} data are qualitatively reproduced.  At $\Phi_{35}=0.17$,
the hard emission shows two main peaks (`A' and `B' in
Figure~\ref{fold}), and a third, lower peak (`C' in
Figure~\ref{fold}). This situation is close to that modeled by Scott
et al. (2000) at similar $\Phi_{35}$, during the progressive
occultation of leading and training peaks of the hard beam. At
$\Phi_{35} \sim 0.27$, when the main components are occulted, we only
observe the survival of a broad, underlying modulation that is
attributed to the magneto-spheric emission. Since this component is
emitted from a larger region at some distance from the neutron star,
it is naturally expected to have a lower modulation as well as a broad
maximum (Figure~\ref{fold}).

The pulse profile close to the short-on is also similar to that
presented by Scott et~al. (2000) at $\Phi_{35}\approx 0.58$ (see their
Figure~10b).  In the EPIC data, we can in fact recognize a main peak
`D' as well a less prominent peak `E' (Figure~\ref{fold}).  A
cross-correlation between different {\sl XMM-Newton} datasets does not
allow a proper phase alignment between main-on and short-on peaks
based on the extrapolation of the pulse timing ephemeris. However,
because the peak `E' is the hardest feature, spectral considerations
suggest that this maximum is associated with the small hard peak and
the feature `D' with the soft peak discussed by Scott
et~al.~(2000). If this is the case, `E' is actually due to direct
emission from the pencil beam, while `D' is the radiation redirected
into the fan beam from the antipodal accretion column.

\subsection{The phase shift between the soft and hard light curves}
\label{shift}

Given the complexity of the source, pulse-phase spectroscopy is of
paramount importance to separate the different spectral components
observed in Her X-1. Using $Einstein$ (McCray et~al. 1982) and
$BeppoSax$ data (Oosterbroek et~al. 1997, 2000) it has been shown
that, during the main-on state, the maximum of the thermal component
and the power law components are shifted by $\sim 250^\circ$ and that
the maximum of the unresolved feature at $\sim 1$~keV is in phase with
that of the blackbody component. The situation is less consistent as
far as the 6.4~keV Fe K line is concerned: Choi et~al. (1994) have
shown that its intensity is modulated in phase with the soft emission,
suggesting a common origin for the two Fe lines, while Oosterbroek et
al. (2000) have found it correlated with the hard (power law)
emission.

The shift in phase between hard and soft emission can be explained if
the latter results from re-processing of hard X-rays in the inner part
of the accretion disk. If a non-tilted disk intercepts (and
re-processes) a substantial fraction of the hard beam from the neutron
star, the expected phase difference between direct and reflected
component is $180^\circ$.  Therefore, the value of $\sim 250^\circ$
determined using $Einstein$ and $BeppoSax$ data has been associated
with the disk having a tilt angle. If the tilt of the disk changes
with phase along the 35~d cycle (as predicted by the precessing disk
models, see Gerend \& Boyndon 1976) the shift in phase should
therefore vary with $\Phi_{35}$. However, both Einstein and Sax data
were obtained at the same $\Phi_{35}$, i.e. during the main-on state
($\Phi_{35} =0.1$ for $Einstein$; $\Phi_{35} =0.07-0.15$ for
$BeppoSax$ in 1997 and $\Phi_{35} =0.1-0.2$ for $BeppoSax$ in
2000). Oosterbroek et~al. (2000) also observed the source at
$\Phi_{35} =0.5$ and found that the pulse phase difference in the
short-on and main-on state are consistent.  This is not surprising,
since symmetry considerations allow for the same behaviour at
$\Phi_{35} =0.0$ and 0.5. A tracking of the phase difference between
the two components over the entire cycle was therefore required.

Here we have found that not only the phase shift derived from {\sl
XMM-Newton} main-on data is considerably different from previous
observations made in the main-on, but it continues to change
dramatically during the other
two observations. This suggests that we are observing, for the first
time, a {\it substantial and continuous variation in the tilt of the
disk}, which is what we would expect from a system which had a
precessing accretion disc. It should be noted that the interpretation
of the phase shift observed at the short-on may be affected by a
systematic error, depending on whether during the observation the soft
peak `D' is higher than the small hard peak `E' or vice-versa (see
\S \ref{enres}).

\subsection{The fluorescent line variation}
\label{line}

To investigate in more detail the possible common origin of the soft
component and the Fe line at 6.4~keV, we have derived the line
parameters as a function of the spin phase, $\phi_{spin}$ (see \S
\ref{pulsespec} and Figure~\ref{fitph}).  At $\Phi_{35}$=0.26 and
0.60, there is little evidence for a significant variation in the line
parameters.

The observation made during the main-on ($\Phi_{35}=0.17$) clearly
shows that the soft flux below 0.7~keV and the equivalent width all
exhibit a common minimum at 0.2\ltae$\phi_{spin}$\ltae0.4, which, in
turn, is shifted with respect to that of the hard emission. This
supports the idea that the 6.4~keV Fe line originates from
fluorescence from the relatively cold matter of the illuminated spot
where the soft emission is reprocessed.  In this case, the flux
emitted in the line may provide a lower limit to the size of the spot
because the observed line energy allows us to put an upper limit to
the ionization degree and to the temperature of the emitting region,
$T$\ltae 0.3~keV (Kallman \& McCray 1982). This is in principle
appealing since it allows us to constrain the size of the illuminated
spot in a way which is less subject to absorption (compared with the
usual methods based on the value of the soft flux).  However, assuming
a spherical spot and a distance to the source of 6.6~kpc (Reynolds
et~al. 1997), the inferred lower limit for the radius is $R$ \gtae
0.5~km, too low to add any significant constraints. For comparison,
assuming the same distance for the source, the radius of the
equivalent blackbody derived from the soft flux varies between $\sim
140$~km and 500~km.

We have also found evidence for a variation in the Fe line parameters
over the 35~d period. The line flux and the line width is greatest at
the main-on state. We also show that the line energy is significantly
higher in the main-on state and lowest in the low state. In the
main-on state, the energy of the Fe line (6.540keV) corresponds to Fe
XX-XXI, while in the low state (6.385keV) it is from low ionisation
species.

This suggests two possible explanations for both the line broadening
and the centroid displacement: 1) an array of Fe K fluorescence lines
exists for a variety of charge states of Fe (anything from Fe I-Fe
XIII to Fe XXIII); 2) Comptonization from a hot corona with a
significant optical depth for a narrower range of charge states
centered around Fe XX. Similar line broadening have been observed in
some Low Mass X-Ray binaries observed using $ASCA$ (Asai et~al. 2000).

The Fe line broadening may also be explained in terms of Keplerian
motion, if the inner disk (or some inner region) comes into view
during the main-on state. If this is the case, at $\Phi_{35}=0.17$ the
Keplerian velocity will be $\sim 13000$ km/sec. This, in turn,
corresponds to a radial distance of $\sim 4 \times 10^8$~cm (for a
neutron star of 1.4\Msun), which is close to the magneto-spheric
radius for a magnetic field of $\sim 10^{12}$~G.

\section{Acknowledgments}

We would like to thank both Tim Oosterbroek and Uwe Lamers for
providing help in resolving the timing problems found in the EPIC pn
data. This paper makes use of quick-look results provided by the
ASM/RXTE team whom we thank. We are grateful to Mariano Mendez, Vadim
Burwitz and Mark Cropper for useful discussions.

\end{document}